\documentclass[aps,prl,reprint,superscriptaddress,showpacs]{revtex4-1}

\usepackage{graphicx}

\begin{document}


\title{Magnetic-field induced multiferroicity in a quantum critical frustrated spin liquid}


\author{F. Schrettle}
\author{S. Krohns}
\author{P. Lunkenheimer}
\email[Corresponding author. ]{Peter.Lunkenheimer@Physik.Uni-Augsburg.de}
\author{A. Loidl}
\affiliation{Experimental Physics V, Center for Electronic Correlations and Magnetism, University of Augsburg, 86135 Augsburg, Germany}
\author{E. Wulf}
\author{T. Yankova}
\altaffiliation{Permanent address: Chemistry Department, M. V. Lomonosov
Moscow State University, Moscow, Russia}
\author{A. Zheludev}
\affiliation{Neutron Scattering and Magnetism Group, Laboratorium f\"{u}r Festk\"{o}rperphysik, ETH H\"{o}nggerberg, 8093 Z\"{u}rich, Switzerland}



\begin{abstract}
Dielectric spectroscopy is used to check for the onset of polar order in the quasi one-dimensional quantum spin system Sul-Cu$_2$Cl$_4$ when passing from the spin-liquid state into the ordered spiral phase in an external magnetic field. We find clear evidence for multiferroicity in this material and treat in detail its $H$-$T$ phase diagram close to the quantum-critical regime.
\end{abstract}




\maketitle

Spin-driven ferroelectrics have turned out to be cornerstone model systems for multiferroics. After the first experimental realization in multiferroic manganites \cite{Kimura2003}, the simultaneous electrical and magnetic ordering in these systems has been explained by spin currents \cite{Katsura2005} or by an inverse Dzyaloshinskii-Moriya interaction \cite{Sergienko2006,*Mostovoy2006}. In both cases a non-collinear helical spin arrangement, which breaks the inversion symmetry between neighboring spins, is responsible for the generation of a finite polarization obeying strict symmetry constraints with respect to the spin structure. In many quasi low-dimensional spin systems, specifically in Heisenberg spin chains and ladders, the necessary complex non-collinear spin structures are often stabilized and enhanced by strong quantum spin fluctuations. Indeed ferroelectricity has been detected in some antiferromagnetic quantum spin chains, e.g., in LiCu$_2$O$_2$ \cite{Park2007} or LiCuVO$_4$ \cite{Naito2007,Schrettle2008}.

In a large class of low-dimensional spin systems, the so-called spin liquids, long range magnetic order is destroyed by zero-point quantum spin fluctuations. It would seem that such materials are poor candidates for the spin-driven ferroelectric effect, as they are not even magnetic. Yet, spontaneous magnetic order can be \textit{induced} in spin liquids through the application of an external magnetic field. This type of transition in specific systems can be described as a Bose-Einstein condensation of magnons \cite{Giamarchi1999,*Rice2002,*Giamarchi2008} and has recently attracted a great deal of attention. In particular, it was extensively studied in a number of prototypical spin ladder materials \cite{Garlea2007,Zheludev2007,Klanjsek2008,Schmidiger2012}. In the presence of geometric frustration of interactions, the high-field ordered phase may actually be a complex helimagnetic structure, susceptible to reverse Dzyaloshinskii-Moriya coupling to the crystal lattice. It stands to reason that ferroelectricity will then emerge at a magnetic-field induced quantum phase transition from a paraelectric spin-liquid state. In the present paper we report the observation of this remarkable phenomenon in a frustrated quantum spin ladder material.

As discussed in detail in Refs. \cite{Garlea2008,Garlea2009,Zheludev2009}, Sul-Cu$_2$Cl$_4$ with the chemical formula Cu$_2$Cl$_4$-H$_8$C$_4$SO$_2$ represents a spin $S=1/2$ four-leg Heisenberg spin tube with a high degree of geometric frustration. In this system long-range magnetic order cannot be established even at the lowest temperatures and the ground state is a spin liquid, characterized by activated spin susceptibility and specific heat.
It is protected from the lowest excited state by a small energy gap ($\Delta=0.52$~meV). This gap can be overcome by a moderate external field that restores long range order beyond a critical field H$_c$~$\approx$~4~T.
Neutron diffraction has shown that the high-field and low-temperature ordered state of Sul-Cu$_2$Cl$_4$ is an incommensurate magnetic phase with a planar helimagnetic structure \cite{Garlea2009,Zheludev2009}. Of special interest is the presence of a field-induced quantum-critical (QC) point in this system, which, moreover, shows unusual values of the order-parameter critical exponents \cite{Garlea2009}. In the present work, we study the dielectric properties of Sul-Cu$_2$Cl$_4$ when passing from the spin liquid into the helimagnetic state. We find strong experimental evidence for ferroelectricity \cite{Khomskii_priv} and, thus, multiferroicity in this material. Hence, this system combines two properties that are in the focus of current research: multiferroicity and a quantum phase transition. We study its electric properties across the extended H-T phase diagram close to the QC point.

Samples of Sul-Cu$_2$Cl$_4$ were grown from solution as described in detail in \cite{Yankova2012}. Before the measurements, the samples were washed in ethanol to clean the surfaces. The crystals are irregularly shaped, oblong platelets with faces set up by the $a$ and $c$ axes, the longer dimension being the chain axis $c$. For the dielectric experiments, the crystals were brought between two electrodes, which were then pressed onto the sample surface. During the measurements, the external magnetic field was aligned along the crystallographic $a$ direction. The dielectric properties were measured at 1 and 10~kHz employing a high-precision capacitance bridge (Andeen-Hagerling AH27000A). For measurements between 1.5~K and 10~K in external magnetic fields up to 14~T, an Oxford cryostat equipped with a superconducting magnet was used. Due to the ill defined sample geometry, the absolute values of the dielectric constant have high uncertainties and, hence, we only provide the capacitances. Since the samples deteriorate over time when exposed to air, several crystals were used for the experiments. To compare the result from the different samples, scaled plots of the capacitance are shown. The presented loss-tangent data are geometry-independent and thus absolute values can be provided.

In the high-field phase, the spontaneous helimagnetic component of magnetization is confined to a plane perpendicular to the applied field and corresponds to a propagation vector $\vec{Q}$ = (-0.22, 0, 0.48) \cite{Garlea2009}. Given this magnetic structure, an external magnetic field $H \parallel a$ induces a spin spiral with the spiral axis $\vec{e} \parallel H \parallel a$ and the propagation vector $\vec{Q}$ close to the (-1,~0,~2) direction. In spin-driven multiferroics, macroscopic ferroelectric polarization $P$ will appear along $\vec{e} \times \vec{Q}$ and, hence, in Sul-Cu$_2$Cl$_4$ we expect to find macroscopic polarization $P$ for magnetic fields larger $H_c$ close to (0,~-2,~0). Therefore for the dielectric measurements a contact configuration was chosen that leads to an ac electric field perpendicular to the $a-c$ plane.

 \begin{figure}[h]
 \includegraphics[width=0.85\linewidth]{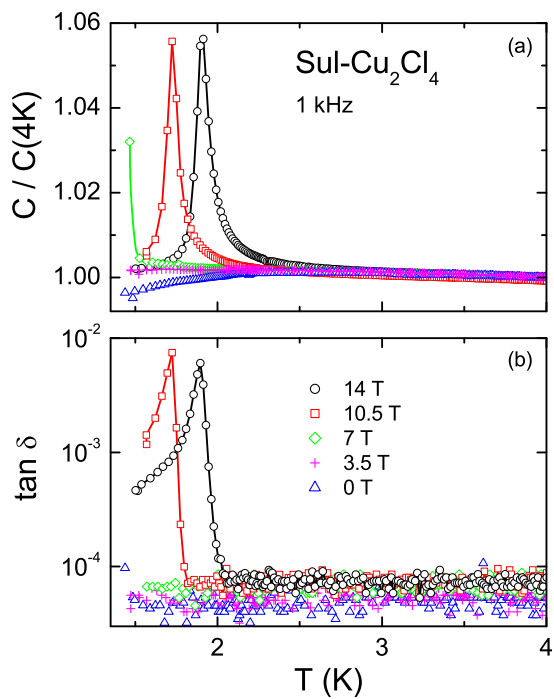}
 \caption{\label{TDep} (Color online) Capacitance (scaled to the value at 4~K) (a) and loss tangent (b) \textit{vs}. temperature of Sul-Cu$_2$Cl$_4$ at a measurement frequency of 1~kHz for various external magnetic fields. The lines are guides to the eyes.}
 \end{figure}

Figure \ref{TDep}(a) shows the capacitance $C$ of Sul-Cu$_2$Cl$_4$, which is a direct measure of the polar susceptibility, as a function of temperature, measured at a frequency of 1~kHz in various external magnetic fields. For 0~T and 3.5~T the capacitance shows no anomalies and is nearly temperature independent. In zero external magnetic field the capacitance shows a crossover from a slight decrease for temperatures below about 2.5~K to a nearly constant behavior at higher temperatures. At 3.5~T, which is close to quantum criticality, $C(T)$ shown in Fig. \ref{TDep}(a) remains practically constant down to the lowest temperatures accessed in the present experiment. At 7~T the capacitance starts to increase significantly below about 1.5~K, signaling the closeness to a phase transition. At 10.5~T a well defined anomaly appears at 1.7~K which shifts to higher temperatures in further increasing magnetic fields. This anomaly signals the transition from the paramagnetic spin-liquid phase into the helimagnetic state \cite{Fujisawa2005,Garlea2009}. It strongly points to the occurrence of ferroelectric polarization, arising simultaneously with the magnetic transition. Similar anomalies have been detected in spin-driven multiferroics, specifically also in the $S = \frac{1}{2}$ chain cuprate LiCuVO$_{4}$ \cite{Schrettle2008}. In conventional ferroelectrics a rather gradual increase at the flanks of the $C(T)$ peak, following a Curie-Weiss law, is expected. In contrast, the very narrow shape of the anomaly observed here and the obvious absence of critical fluctuations signal the improper, field-induced type of the ferroelectric phase transition.

In Fig. \ref{TDep}(b) the loss tangent $\tan \delta(T)$ is shown. For magnetic fields up to 7~T a strongly scattering, very small constant loss is found corresponding to the resolution limits of the device. However, for higher fields the loss rises above background and shows well-defined peaks close to the phase transition. Similar behavior was also found for LiCuVO$_4$ \cite{Schrettle2008}. Neither for $C$ nor $\tan \delta$ we found any significant frequency dependence. This agrees with the expectations for an improper ferroelectric where long-range polar order is induced by the onset of spiral spin order and no slowing down of polar relaxations should occur. Moreover, the observed loss peaks are asymmetric and show very steep high-temperature flanks. Again, this finding is in accord with an improper ferroelectric state, triggered by the helical magnetic order. The significant, frequency-independent enhancement of the loss indicates strong fluctuations of the ordering dipoles close to the phase transition, with a broad distribution of relaxation times.


 \begin{figure}[h]
 \includegraphics[width=0.85\linewidth]{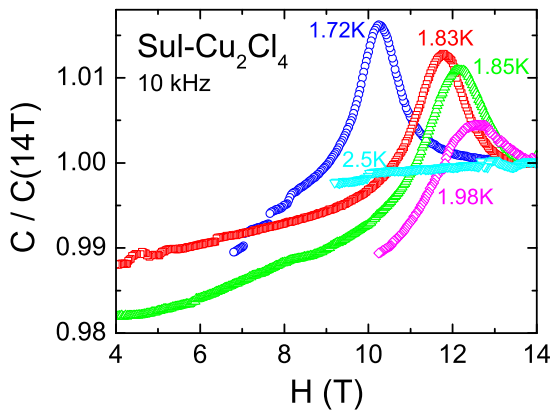}
 \caption{\label{HDep} (Color online) Capacitance (scaled to the value at 14~T) \textit{vs}. external magnetic field of Sul-Cu$_2$Cl$_4$ at a measurement frequency of 10 kHz for several temperatures.}
 \end{figure}

The dielectric response of Sul-Cu$_2$Cl$_4$ as a function of the external magnetic field is shown in Fig. \ref{HDep}, measured at 10~kHz and different temperatures between 1.72 and 2.5~K. The measurements were taken with the same configuration of electric and magnetic field as the temperature-dependent curves in Fig. \ref{TDep}. At 2.5~K and fields up to 14~T the capacitance does not show any significant field dependence. For a temperature of 1.98~K and below, however, a clear maximum is revealed. It shifts towards lower fields with decreasing temperatures. These peaks are complementary to the ones detected in the temperature-dependent measurements shown in Fig. \ref{TDep}. In both cases, the anomalies appear at the transition between the paramagnetic and paraelectric groundstate and the helimagnetic and polar high-field phase. These findings strongly point to multiferroicity induced by an external magnetic field in this material.

\begin{figure}[h]
 \includegraphics[width=0.85\linewidth]{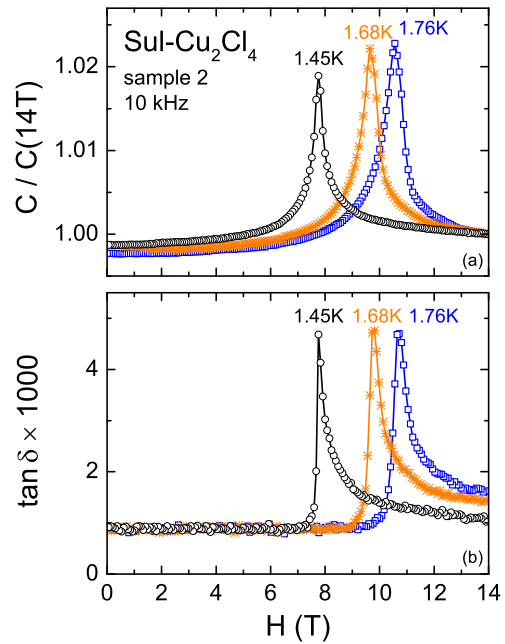}
 \caption{\label{HDep2nd} (Color online) (a) Capacitance (scaled to the value at 14~T) \textit{vs}. external magnetic field of Sul-Cu$_2$Cl$_4$ (second sample) at a measurement frequency of 10 kHz for several temperatures. Frame (b) shows the corresponding loss-tangent curves.}
 \end{figure}

At fields below the onset of the capacitance peaks, $C(H)$ in Fig. \ref{HDep} still shows a significant field dependence, which is weaker than that at the left flanks of the peaks. The origin of this effect is unclear and one may suspect it to be related to QC behavior. However, such behavior is not found in the results obtained for a second sample from a different batch, shown in Fig. \ref{HDep2nd}(a) and, thus, its significance is limited. In the data of Fig. \ref{HDep2nd}(a), again well-pronounced peaks show up, whose positions are compatible with the results on the first sample. In addition, in Fig. \ref{HDep2nd}(b) the magnetic-field dependent loss tangent is shown. At the transition, peaks show up, consistent with the findings in the temperature-dependent data [Fig. \ref{TDep}(b)]. The low-field flanks of these peaks are nearly vertical, quite in contrast to the much smoother increase of $\tan\delta(H)$ found when approaching the transition from the high-field polar state. As outlined above, these frequency independent loss peaks seem to be in accord with an improper ferroelectric transition. Concerning the significant asymmetry, one plausible explanation could be that domain-wall dynamics is the source of the dielectric loss, which arises when ferroelectric domains are abruptly formed at the critical field. For magnetic fields beyond $H_{c}$, the growing domain size and reduced number of domain walls leads to the gradual decrease of $\tan\delta$ deep in the ordered state. On the other hand, the more symmetric peaks of the dielectric constant [Fig. \ref{HDep2nd}(a)] result from polar fluctuations, which exist in the bulk of the material.

We also have performed additional measurements of the electrical polarization \textit{vs}. the electric field aiming at the detection of polarization hysteresis curves. However, no significant non-linear effects in the ferroelectric phase were found. Most likely the ferroelectric polarization is too low to be detectable and dominated by the paraelectric background. In this context, we want to stress the weakness of the observed anomaly in $C(T)$ (Fig. \ref{TDep}(a)). In Sul-Cu$_2$Cl$_4$ its strength amounts approximately 5~\% of the background dielectric constant only. For comparison, in LiCuVO$_4$ it reaches almost 12~\% \cite{Schrettle2008}.

 \begin{figure}[h]
 \includegraphics[width=1\linewidth]{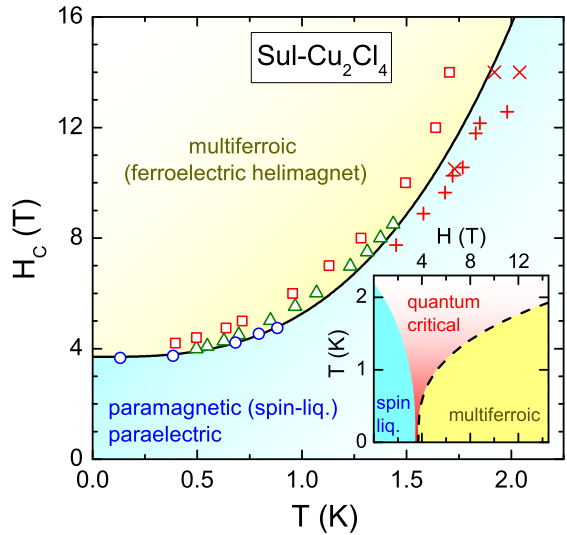}
 \caption{\label{PD} (Color online) Critical field $H_c$ as a function of temperature. The shown data points were obtained from dielectric experiments (present work, $+$: from H dependence, $\times$: from T dependence), neutron-scattering (circles \cite{Garlea2009}), and specific-heat measurements (triangles \cite{Fujisawa2005} and squares \cite{Wulf2011}). The solid line represents a critical power-law \cite{Garlea2009} as described in the text. The inset shows a tentatively proposed QC scenario for the present material with the typical V-shaped QC region. The dashed line corresponds to the same critical law as shown in the main frame.}
 \end{figure}

The present dielectric results allow us to rethink the previously published phase diagram of Sul-Cu$_2$Cl$_4$ \cite{Garlea2009,Wulf2011}. Figure \ref{PD} illustrates the temperature dependence of the critical magnetic field $H_c$, comprising results from neutron-scattering \cite{Garlea2009} (circles) and specific-heat studies \cite{Fujisawa2005,Wulf2011} (triangles and squares). In addition, the results of the temperature- and magnetic-field-dependent measurements of the present work (Figs. \ref{TDep}, \ref{HDep}, and \ref{HDep2nd}) are included ($+$ and $\times$). The phase-boundary line in Fig. \ref{PD} corresponds to the function $H_c(T) = H_{c0} + 1.565 \times T^{1/\nu}$ with a critical field at 0~K of $H_{c0}=3.7$~T and a critical exponent $\nu = 0.34$ as used in ref. \cite{Garlea2009} to describe the data below 1.5~K, treated in that work (circles and triangles). The extrapolation of this function to higher temperatures provides a reasonable description of the additional data shown in Fig. \ref{PD}. The variation of the data points around this line, especially observed at the higher temperatures, indicate a sample dependent scattering of the transitions temperatures and fields, most likely mirroring the experimental difficulties that arise from the rapid deterioration of the samples. The unusual value of the exponent $\nu = 0.34$ and other critical indexes, which are quite distinct from mean field values expected for a Bose-Einstein
condensation of magnons, were previously attributed to either the
chiral nature of the ordered state or a dimensional crossover
phenomenon \cite{Garlea2009}. Our discovery of ferroelectricity of the ordered
state suggests another intriguing explanation: The transition may,
in fact, be in an entirely different universality class from Bose-Einstein condensation, as
it involves a joint magnetoelectric order parameter.

It is well known that close to a QC point, materials show unique properties, whose experimental characterization and theoretical explanation is a very active field of current research \cite{Sachdev2000,*Coleman2005,*Gegenwart2008}. This happens in a "V-shaped" region in the phase diagram, the tip of the "V" being centered at the QC point at zero K \cite{Sachdev2000,*Coleman2005,*Gegenwart2008}. In the inset of Fig. \ref{PD}, the red shaded area  shows this "V-shaped" QC region as tentatively proposed for the present case. Here the dashed line corresponds to the critical law \cite{Garlea2009} as already provided in the main frame of Fig. \ref{PD}. Interestingly, the gradual increase of $C(H)$ observed in the magnetic-field sweeps at low fields (Fig. \ref{HDep}) arises in the V-shaped area and, thus, may be suspected to be a signature of the QC regime. However, as mentioned above this feature could not be reproduced in a second sample and is of limited significance. Overall, the inset of Fig. \ref{PD} provides a plausible scenario, which needs to be corroborated by further measurements, especially at lower temperatures, looking for the dielectric signature of the quantum criticality. However, one should be aware that magnetic-field dependent dielectric measurements at these sub-He temperatures are a non-trivial task, which is further hampered by the rapidly deteriorating samples. In addition, at present there is practically no theoretical guidance available to understand quantum phase transitions and criticality in ferroelectric systems. Thus, we hope that the present study will stimulate both experimental and theoretical research in that direction.

In summary, significant anomalies were found in both temperature- and magnetic-field-dependent measurements of the dielectric properties of Sul-Cu$_2$Cl$_4$. They provide strong evidence for polar order and, thus, for multiferroicity in this compound and comply with the symmetry constraints for ferroelectricity driven by helimagnetic spin order. Just as for the magnetic degrees of freedom, the electrical ordering is only observed under the application of external magnetic fields and is suppressed in zero field. To clarify the role of quantum criticality for the dielectric and other properties of this material, further investigations are necessary.

\begin{acknowledgments}

This work was supported by the Deutsche Forschungsgemeinschaft via the Transregional Collaborative Research Center TRR 80 and partly via the Research Unit FOR960. Work at ETHZ was supported by the Swiss National Science Foundation through Division II and MaNEP.

\end{acknowledgments}


\bibliography{SulCu2Cl4BIB}

\end{document}